\documentclass[12pt]{article}
\usepackage[margin=0.95 in]{geometry}
\usepackage{amsmath}
\usepackage{amssymb,amsfonts}
\usepackage[all]{xy}
\usepackage{graphicx}
\usepackage[utf8x]{inputenc}
\usepackage{amsmath}
\usepackage{amssymb}
\usepackage{float}
\usepackage{array}
\usepackage{tikz}
\usepackage{mathtools}
\usepackage{mathrsfs}
\usepackage{hyperref}
\usepackage{cite}
\numberwithin{equation}{section}
\numberwithin{equation}{section}
\numberwithin{table}{section}
\begin{document}

\thispagestyle{empty}

\vspace*{3cm}
{}

\noindent
{\LARGE \bf Twisted Massive Non-Compact Models }
\vskip .4cm
\noindent
\linethickness{.06cm}
\line(10,0){447}
\vskip 1.1cm
\noindent
\noindent
{\large \bf Songyuan Li and Jan Troost}
\vskip 0.25cm
{\em
\noindent
Laboratoire de Physique Th\'eorique
de l'\'Ecole Normale Sup\'erieure \\
\hskip -.05cm
 CNRS,
 PSL Research University and Sorbonne Universit\'e, Paris, France
}
\vskip 1.2cm

\vskip0cm

\noindent {\sc Abstract: } {We study interacting massive $N=(2,2)$ supersymmetric field
theories in two dimensions which arise from deforming conformal field theories with a continuous spectrum. Firstly,
we deform $N=2$ superconformal Liouville theory with relevant operators, and twist the theory into a topological 
quantum field theory. These theories can be thought of as twisted generalized Landau-Ginzburg models with negative power superpotential.
We determine the structure constants of the chiral ring and therefore all correlators of these topological quantum field theories.  
We provide general formulas for the deformed operators of given charge  as well as explicit solutions to the WDVV equations. Secondly, 
we  analyze the topological anti-topological sector of the theory.
We compute the metric at the conformal point through localization. Moreover, we show that the
topological-anti-topological fusion differential equations on the metric in a family of non-compact theories
takes the affine Toda form. The metric as a function of the family of theories is identical to the metric in certain deformed
compact models. For a negative cubic power  superpotential, for instance, it is governed by the Painlev\'e III differential equation.

\vskip 1cm

\pagebreak

\newpage
\setcounter{tocdepth}{2}
\tableofcontents

\section{Introduction}
Over the last seven decades, our understanding of quantum field theory has
steadily progressed. A crucial consequence is that quantum field theory has become the paramount tool in our description of the fundamental interactions and a wealth of
other natural phenomena. Nevertheless, we are still far from a complete understanding of quantum field theories, and even in the last decades, these theories have offered 
up interesting theoretical puzzles and breakthroughs.

A modern technique in the study of quantum field theory is the topological twist
of supersymmetric theories \cite{Witten:1988ze}. The twist has had wide physical
as well as mathematical applications. One such application was the understanding of the topological twisting of
conformal as well as massive compact $N=(2,2)$ supersymmetric field theories
in $1+1$ dimensions with discrete spectrum (see e.g. \cite{Eguchi:1990vz,Martinec:1988zu,Vafa:1988uu,Witten:1993jg}).  That effort was motivated through various
trains of thought. For instance, the identification of infrared fixed
points of ultraviolet Lagrangians, and the classification of
such fixed points is a contribution to our global map of the space of quantum field
theories. This is equally true of the study of the properties of massive deformations of known
fixed points. There was also strong
motivation from the application of these quantum field theories as
theories on a superstring worldsheet, after coupling to
two-dimensional supergravity, and the relation  of these string theories \cite{Witten:1989ig,Distler:1989ax,Li:1990ke,Dijkgraaf:1990dj} to matrix models \cite{Brezin:1990rb,Douglas:1989ve,Gross:1989vs}.

In this paper, we set out to generalize the analysis of
topological twists of $N=(2,2)$ supersymmetric two-dimensional
theories to theories with a non-compact target space. In other words, we study topological quantum field theories that arise upon
massive deformation of two-dimensional interacting superconformal
field theories with continuous spectrum. After twisting, we determine the spectrum, the chiral ring, and therefore, the correlation functions on any Riemann surface.
We moreover study the topological-anti-topological sector  \cite{Cecotti:1991me}, namely the chiral-anti-chiral metric of the theory 
at the fixed point, as well as the dependence of the topological-anti-topological metric on a coupling constant.

There have been studies of the relation between generalized Landau-Ginzburg models with negative power monomial superpotential and 
non-linear sigma-models, as well as the string theory that results from them when coupling to gravity. These have
given precious information about the model at level one coupled to gravity \cite{Ghoshal:1995wm,Ghoshal:1993qt,Hanany:1994fi},
as well as the relation of the monomial at generic level  to string compactifications 
\cite{Ooguri:1995wj,Hori:2001ax,Eguchi:2000tc,Eguchi:2004ik,Ashok:2007ui} and to $N=2$ Liouville theory  through a comparison of their
elliptic genera \cite{Troost:2010ud}. The topological twist of the massive deformations of these models is  unexplored territory.

In section \ref{TCFT} we discuss the correlators of a topologically twisted superconformal $N=2$ Liouville
theory. We deform the theory by relevant operators and discuss the properties of the resulting topologically
twisted massive non-compact theories in section \ref{TQFT}. In section \ref{conformalmetric} we determine
the metric or non-chiral structure constants at the conformal point and how to compute them using
localization. We then analyze the dependence of the chiral-anti-chiral metric on one parameter
using the $tt^\ast$ equation in section \ref{ttstar}. Throughout, we very briefly recall compact counterparts for the analyses we perform in the non-compact model (though the reader
will need to refer back to the original literature for a detailed treatment of the compact models).
Thus, we are able to highlight useful analogies as well as crucial differences with the class of non-compact models which are the subject of our paper. We conclude with
a discussion of wider applications and future directions in section \ref{conclusions}.

\section{The Topological Conformal Field Theory}
\label{TCFT}

In this section, we describe topological conformal field theories. Firstly, we
recall the topological conformal field theories that arise from twisting the $N=2$ minimal model conformal field theories  \cite{Eguchi:1990vz,Martinec:1988zu,Vafa:1988uu,Witten:1993jg}.
Secondly, we introduce  non-compact counterparts (see e.g. \cite{Eguchi:2000tc,Eguchi:2004ik,Eguchi:2004yi,Israel:2004ir,Ashok:2007ui,Troost:2010ud} for necessary background).

\subsection{The Topological Compact Model}
\label{compactTCFT}
Quantum field theories in two dimensions with $N=(2,2)$ supersymmetry are sufficiently constrained to frequently allow for the identification of their
interacting infrared fixed points.
 An
 example of such a succesful identification is the proposal that the
 $N=2$ minimal model conformal field theory is the infrared fixed point
 of a $N=2$ Landau-Ginzburg model with as field content an $N=(2,2)$ chiral
 superfield $\Phi$, and interactions dictated by the superpotential $W=\Phi^{k_c}/k_c$. The positive
 integer power $k_c$ is related to the central charge of the $N=2$
 minimal model by the relation $c=3-6/k_c$. We will refer to the integer $k_c$ as the level of the model.\footnote{We denote the level of the compact model
by $k_c$, to contrast with the level  of the non-compact model which will be denoted $k$.} 
 Considerable evidence for
 this identification was amassed over the years
\cite{Martinec:1988zu,Vafa:1988uu,Witten:1993jg}. 

Part of the evidence lay in the understanding of the topological subsector of the infrared fixed point, the topological conformal field theory \cite{Eguchi:1990vz}.
We briefly review the properties of this topologically twisted compact $N=2$ superconformal field theory. Our starting point is the $N=2$ minimal model, which can  be thought of as a supersymmetric 
sigma-model on the coset $SU(2)/U(1)$. We concentrate on the A-type diagonal modular invariant.
We moreover focus on the (chiral,chiral) ring \cite{Lerche:1989uy} of the  minimal model at central charge $c=3-6/k_c$. It is made up of the operators in the spectrum whose conformal dimension is half their R-charge, both for the left- and the right-movers. In the NSNS sector, they can be identified as arising from the parent $SU(2)$
operators with spin $j$ which have angular momentum $m$ equal to their spin, $j=m$. The spin $j$ runs over the values $j=0,1/2,\dots,(k_c-2)/2$ where $k_c-2$ is the level of the bosonic
$SU(2)$ current algebra of the parent theory on the group $SU(2)$. The chiral ring thus has $k_c-1$ elements that have R-charges in the set $\{ 0,1/k_c,2/k_c, \dots, 1-2/k_c \}$ in the NSNS sector. 
The ring structure constants agree with those of the polynomial ring $\mathbb{C}[X]$ modded out by the ideal generated by $X^{k_c-1}$, where $X$ has R-charge $1/k_c$. 
The ring has a unit. In the Ramond-Ramond sector, after spectral flow by half a unit in the appropriate direction, the chiral operators map to ground states. 
These have R-charges shifted down by $c/6=1/2-1/k_c$ and therefore
take values in the set $\{ -1/2+1/k_c,-1/2+2/k_c, \dots, 1/2-1/k_c \}$ which runs from $-c/6$ to $+c/6$ with increments of $1/k_c$. 
The topological metric, defined to be the expectation value of two chiral operators, which we can identify with $X^i$ and $X^j$, in the topologically twisted theory on the sphere
equals (see e.g. \cite{Dijkgraaf:1990dj})
\begin{equation}\eta_{ij}
 =\langle X^i X^j \rangle_0
 =  \delta_{i+j,k_c-2} \, .
\end{equation}
We note that the vacuum in this compact theory is normalizable, and that the vacuum state, with conformal dimension and R-charge both equal to zero, survives
the topological twist.
The structure constants of the ring are related to the three-point functions via the topological metric, and they equal
\begin{eqnarray}
{c_{ij}}^l &=& \delta_{i+j,l}
\nonumber \\
c_{ijl} &=& \langle X^i X^j X^l \rangle_0 = {c_{ij}}^m \eta_{ml}=\delta_{i+j+l,k_c-2} \, ,
\end{eqnarray}
where $i,j,l$ need to be in the spectrum in order for the structure constants or three-point function to be non-zero. This information is sufficient to describe all
correlators of the topological conformal field theory on any Riemann surface, by cutting and sewing.
The topological conformal field theory has an alternative description in terms of a twisted $N=2$ Landau-Ginzburg model. The Landau-Ginzburg theory has a superpotential
\begin{equation}
W = \frac{X^{k_c}}{k_c}
\end{equation}
and the chiral ring is again the polynomial ring $\mathbb{C}[X]$ modded out by $W_X=\frac{dW}{dX}=X^{k_c-1}$. 

Establishing the dictionary for  these supersymmetric quantum field theories
took several years, and ingenious checks on the identification of the fixed points were performed. 
We refer to the extensive literature for   detailed discussions. Our lightning review mostly serves as a point of reference for the models to come,
which are of an entirely different nature, yet show remarkable resemblances.

\subsection{The Topological Non-Compact Model}
\label{noncompactTCFT}
More than twenty years ago, it was suggested that the correspondence
between Landau-Ginzburg models and conformal field theories could be
usefully extended to non-compact models, with continuous spectrum and
a central charge $c>3$ \cite{Ooguri:1995wj}. See also
\cite{Ghoshal:1995wm,Ghoshal:1993qt,Hanany:1994fi}. 
 The $N=2$ Liouville theory can indeed be understood as a
linear dilaton theory, with an exponential supersymmetric  potential
which is marginal. Importantly, the linear dilaton profile fixes part of
the asymptotics \cite{Hori:2001ax} and renders the superpotential term consistent with conformal symmetry. 
These are two aspects that differ from the compact theories.
It was also shown that the $N=2$ Liouville theory
is T-dual to the $N=2$ coset conformal field theory on the cigar
$SL(2,\mathbb{R})/U(1)$ \cite{Hori:2001ax,Israel:2004jt}.  Earlier on, a more rudimentary probe of the
$N=2$ Liouville theory, namely the Witten index (defined as a periodic path integral on the torus), was computed in
\cite{Girardello:1990sh}, and demonstrated to depend on the choice of
asymptotic radius. Given an $N=2$ Liouville theory of central charge
$c=3 + 6/k$, where $k$ is a positive integer, the natural choices of
asymptotic radii are multiples of $\sqrt{\alpha'/k}$. The Witten index is equal to the chosen multiple.  Since then, our understanding of the
spectrum of the $SL(2,\mathbb{R})/U(1)$ conformal field theory has
progressed both through an analysis of the bosonic partition function
\cite{Hanany:2002ev} and its supersymmetric counterpart \cite{Eguchi:2004yi,Israel:2004ir}, as well as
through an analysis of the elliptic genus \cite{Troost:2010ud}.

Our topological non-compact model of choice will be a topological twisted $N=2$ Liouville theory at central charge $c=3+6/k$
and with an asymptotic radius equal to $\sqrt{k \alpha'}$.  We choose the level $k$ to be a positive integer. We make this choice of radius since we want the number of ground states to be equal to the level (minus one). The theory is a $N=(2,2)$  theory with chiral superfield $\Phi$ and superpotential
\begin{equation}
W = \mu \exp ( b \Phi) \label{superpot}
\end{equation}
where $b=\sqrt{k/2}$, and the theory is supplemented with a linear dilaton at infinity of slope $1/b$.
Establishing the chiral ring of the theory is subtle. (See also \cite{Ashok:2007ui} for a discussion.) The Witten index calculation  \cite{Girardello:1990sh}
cleverly cancels a volume divergence (as detailed in \cite{Troost:2010ud}), and gives a result equal to $k$ (namely, the number of coverings of the minimal circle
radius $\sqrt{\alpha'/k}$). The potential Ramond-Ramond ground states are identified in the NSNS (chiral,chiral)
ring of the conformal field theory as follows. The $N=2$ Liouville theory under consideration is T-dual to a cigar coset superconformal
field theory on the coset $SL(2,\mathbb{R})_k/U(1)$ \cite{Hori:2001ax,Israel:2004jt}, modded out by $\mathbb{Z}_k$. In the coset conformal field theory we can identify the 
(chiral,chiral) states of the original Liouville theory as arising from 
discrete representations $D_j^+$, with lowest weight state with spin component $j$.
The list of possible values for the spin $j$ contains at least the set $j=1,3/2, \dots,(k-1)/2, k/2$. The supplementary values $j=1/2$ and $j=(k+1)/2$ correspond to almost, but not quite
normalizable states, in the following sense. 
The discrete lowest weight representation $D_{1/2}^+$ is sometimes referred to as mock discrete. It does not arise
in the Plancherel formula for the representation space decomposition of quadratically integrable functions on the group manifold $SL(2,\mathbb{R})$. Intuitively, it is analogous to the ground state of conformal
field theories in two dimensions with any non-compact target space, for instance a flat space $\mathbb{R}^n$. The ground state norm is infinite, and proportional to the volume of the target space.  
Similarly, in Liouville theory, the state at the end of the continuous spectrum (namely in the representation $D_{1/2}^+$) is almost, but not quite, normalizable. Spectral flow then
implies that the same type of argument applies to the representation with $j=(k+1)/2$ \cite{Maldacena:2000hw}. The bounds on the spin were first suggested in  
\cite{Giveon:1999px} and firmly established in \cite{Maldacena:2000hw} and \cite{Hanany:1994fi}. 

This subtlety is also manifest after regularizing the conformal field theory
with a linear deformation in $\Phi$, as analyzed in \cite{Girardello:1990sh}:
\begin{equation}
W_{m} = \mu \exp ( b \Phi) -m \Phi \, .
\end{equation}
In \cite{Girardello:1990sh} it is shown that after regularization with the mass term, only $k-1$ normalizable ground states can be identified. Relatedly, in the calculation of the non-compact elliptic genus
\cite{Troost:2010ud} a regularization choice decides on whether the $j=1/2$ (or $j=(k+1)/2$) state contributes to the holomorphic part of the elliptic genus, or not. Thus, the volume divergence
contaminates the ground state counting and even the supersymmetric index must be interpreted with care. See \cite{Troost:2017fpk} for a much more detailed discussion of closely related intricacies.

Despite the subtleties we encounter, we can draw a number of conclusions  at this stage. First of all, the unit operator is not normalizable, and is not in the topological
cohomology. The chiral ring is naturally without unit. Nevertheless, it is true that all rings can be rendered unital by 
adding a unit (operator) by hand. If we do, we must remember that there is no (normalizable) state-operator correspondence for the unit operator.
Secondly, the ring must contain the elements with R-charge $\{ 2/k,3/k, \dots, 1 \}$ which are strictly normalizable. They correspond to the operators
$ \exp(\frac{n}{2b} \Phi)$ where $n=2,3,\dots,k$.  There is another candidate operator at $j=1/2$,
namely $\exp(\frac{1}{2b} \Phi)$ corresponding to $(h,q)=(1/2k,1/k)$, which permits the interpretation as the state in which to evaluate
topological field theory correlators. It is a zero momentum state (since $j=1/2$), much as the $SL(2,\mathbb{C})$ invariant vacuum in a conformal
field theory with target space $\mathbb{R}^p$. Importantly, it does carry a conformal dimension and R-charge. Finally, it has a counterpart at $j=(k+1)/2$
(that arises from spectral flow of a $D^-_{1/2}$ representation in the parent $SL(2,\mathbb{R})$ theory \cite{Maldacena:2000hw}).

To further discuss the operator ring, it is handy to introduce the field theory variable
\begin{equation}
Y^{-1} = e^{\frac{1}{2b} \Phi} \, ,
\end{equation}
in terms of which the superpotential (\ref{superpot}) reads
\begin{equation}
W = \frac{Y^{-k}}{k} \, , \label{superpot2}
\end{equation}
for the choice $\mu = 1/k$. This superpotential has been proposed as a starting point for analysis a while back \cite{Ooguri:1995wj}, but it was mostly made sense
of directly at the conformal fixed point \cite{Hori:2001ax}, in the field variable $\Phi$.
One of the goals in this paper is to show that indeed, the negative power superpotential (\ref{superpot2}) 
leads to a useful and efficient description of the topological model.
The variable $Y$ will be handy, despite having a non-canonical kinetic term, and various other perturbing features.

In these variables then, we will formulate and study two rings. The first ring is the (strict)  chiral ring $R$, given by linear combinations
of the set $\{ Y^{-2},Y^{-3}, \dots, Y^{-k} \}$. We can think of the chiral ring as a subspace of the ring $\mathbb{C}[Y^{-1}]$ of polynomials in $Y^{-1}$.
Moreover, the chiral ring is a subspace of the ideal  $Y^{-2} \mathbb{C}[Y^{-1}]$ which has no unit. In the latter ideal, the chiral ring is the quotient by the ideal $\langle Y^{-k+1} \rangle$ generated
by $Y^{-k+1}=-Y^2 W_Y$. Indeed, note that we generate only powers $Y^{-k-1}$ or higher by multiplying $Y^{-k+1}$ by elements in $Y^{-2} \mathbb{C}[Y^{-1}]$. We can alternatively think of the chiral ring
as being obtained by setting $Y^{-k-1}$ to zero, and we denote these statements as $R=Y^{-2} \mathbb{C}[Y^{-1}]/\langle Y^{-k+1} \rangle=Y^{-2} \mathbb{C}[Y^{-1}]/\{ Y^{-k-1}=0 \}$.
The (strict) chiral ring is made up of all linear combinations of the $k-1$ monomials in the chiral ring that are strictly normalizable.  The Ramond-Ramond sector R-charges of the operators are
$\{  -1/2-1/k, \dots, 1/2-1/k\}$.\footnote{Note that the spectrum of R-charges for the
states that are strictly normalizable is equal for a non-compact model at level
$k$ and a compact model at level $k_c=k$. We will make use of this remark in section \ref{ttstar}.}
The second ring is the extended chiral ring $R_{ext}$, which consists of operators that are linear combinations of
the $k+1$ operators $\{ Y^{-1}, Y^{-2}, \dots , Y^{-k}, Y^{-k-1} \}$ , with a
standard multiplication rule, and such that $Y^{-k-2}$ is equivalent to zero. Again we can describe the extended chiral ring abstractly.
The ring of polynomials has an ideal $Y^{-1} \mathbb{C}[Y^{-1}]$ generated from the monomial $Y^{-1}$. This ideal is a subring without unity.
In the latter ring, we consider the ideal $\langle W_Y \rangle$ generated by $W_Y=-Y^{-k-1}$. We can then consider the quotient ring of $Y^{-1} \, \mathbb{C}[Y^{-1}]$ modded out by the
ideal $\langle W_Y \rangle$. Again, note that the first element we will put to zero is $Y^{-k-2}$, since we multiply elements of the ring into the ideal, and the lowest
order element in the ring is $Y^{-1}$. Thus, we can symbolically write the extended chiral ring as $R_{ext}=Y^{-1} \, \mathbb{C}[Y^{-1}]/\langle W_Y \rangle=Y^{-1} \, \mathbb{C}[Y^{-1}]/\{ Y^{-k-2}=0 \}$.
The R-charges of the extended chiral ring elements in the NS sector are $\{  1/k, 2/k, \dots, 1, 1+1/k
\}$. 
In the Ramond-Ramond sector the R-charges are $\{ -1/2, -1/2-1/k, \dots, 1/2-1/k, +1/2\}$. The set of R-charges  is symmetric under charge conjugation.
In both rings, there is an operator that allows for supersymmetric marginal deformation of R-charge $1$ -- it is the original superpotential term.
 Both the strict and the extended chiral rings will be conceptually useful.

Our next step in defining the non-compact topological conformal field theory is to find a non-zero, non-degenerate symmetric bilinear form $\eta$ on the ring in the guise
of a spherical two-point function for the generating operators $Y^{-i}$.  We wish to define the topological metric $\eta_{ij}$ again as a two-point function
\begin{equation}
\eta_{ij} \overset{?}{=} \langle 0| Y^{-i} Y^{-j} |0 \rangle_0 \, , \label{propexpr}
\end{equation}
but there are several reasons why this is not entirely trivial to make sense of. The first reason is that the state
$|0\rangle$ is the non-normalizable $SL(2,\mathbb{C})$ invariant state, which belongs neither to the original, nor to the twisted theory. 
Therefore, rather than take the expression (\ref{propexpr}) literally, we interpret the expression as referring to the expectation value
\begin{equation}
\eta_{ij} 
= \langle  Y^{-1} | Y^{-i+1} Y^{-j+1} | Y^{-1}  \rangle_0
\, , \label{noncompactmetric2}
\end{equation}
where the state $|Y^{-1} \rangle$  in which we take the vacuum expectation value is the state corresponding to
the operator $Y^{-1}$, namely the almost-normalizable state with conformal dimension and R-charge $(h,q)=(1/2k,1/k)$.
Secondly, we analyze R-charge conservation in the topologically twisted theory.
The charge conservation equation for topological correlators evaluated in the $SL(2,\mathbb{C})$ invariant ground state
is (see e.g. \cite{Dijkgraaf:1990dj})
\begin{equation}
\sum_{i=1}^r q_i + \sum_{i=r+1}^{r+n} (q_i-1) = \frac{c}{3} (1-g) \, 
\end{equation}
where $g$ is the genus of the Riemann surface on which we compute the correlators, the charges $q_{1,\dots,r}$ are the R-charges of the unintegrated vertex operators,
and $q_{r+1,\dots,r+n}$ are the R-charges of the integrated vertex operators. The R-charge at infinity on the right hand side arises from the twisting
of the energy momentum tensor by the derivative of the R-current.
In our context, the R-charge conservation must moreover be modified appropriately by the R-charge of the states in which we take the expectation value (see equation (\ref{noncompactmetric2})).

In practice, for the spherical two-point function we need
\begin{equation}
 q_i+q_j = c/3\, .
\end{equation}
It is thus natural to propose the result 
\begin{equation}
\eta_{ij} = \delta_{i+j,k+2} \, . \label{topY}
\end{equation}
Let us briefly discuss this result. One way to view the result is as an expectation value for the operator $Y^{-k-2}$, which may be surprising since the operator is formally equivalent to zero
in the chiral ring.
 An alternative view on the expectation value is that we evaluate the expectation value of
the operators $Y^{-k}$  in the (almost-normalizable) state corresponding to the operator $Y^{-1}$. Thus, the seemingly surprising 
feature of giving a vacuum expectation value to the operator $Y^{-k-2}$, which is set to zero in the ring under the constraint
$-Y^{-1} W_Y=Y^{-k-2} \equiv 0$ can either be viewed as due to the strong divergence in the norm of the state $|0 \rangle$ or, alternatively,
can be read as corresponding to the vacuum expectation value of (non-zero) operator $Y^{-k}$ in the state generated by $Y^{-1}$. We will provide
more detail to the last interpretation shortly.

Meanwhile, we observe that the metric $\eta_{ij}$ provides a mild cross check on the identification of the extended and strict rings. 
Indeed, the topological metric is a non-degenerate symmetric bilinear form on the extended ring $R_{ext}$ as well as on the vectorial subspace
corresponding to the ring $R$.

Finally, given the metric $\eta$ and charge conservation, it is natural to also propose the three-point function
\begin{equation}
c_{ijl}=  \delta_{i+j+l,k+2} \, .
\end{equation}
We will further argue these proposals in subsection \ref{smalldeformations}, and we will derive interesting consequences of accepting them in section \ref{TQFT}.

\subsection{Small Deformations}
\label{smalldeformations}
In this subsection, we study how the topological conformal field theories behave under a small deformation that gives rise to separated massive
vacua. The analysis provides extra insight into the  features typical of the non-compact model.

\subsubsection{Near the Compact TCFT}
To introduce our tools, we reconsider the topological minimal model correlators from a Landau-Ginzburg perspective.
The correlators of the topological Landau-Ginzburg theory on a genus $g$ Riemann surface are computed through the formula \cite{Vafa:1990mu} 
\begin{equation}
\langle X^{i_1} \dots X^{i_n} \rangle_g = \sum_{W_X=0} X^{i_1} \dots X^{i_n}  H^{g-1} \, ,
\label{LGformula}
\end{equation}
where $H=W_{XX}$ is the Hessian of the superpotential. We compute the correlator by first splitting the classical vacua
by adding a regulator term $\epsilon X$ to the superpotential, and afterwards take the regulator to zero to recover the conformal correlators. 
This is a well-defined procedure in this context, and the correlators agree with those that follow from the chiral ring relations provided in subsection \ref{compactTCFT}.
It is easy to show this explicitly. The calculation includes the steps:
\begin{align}
W&= - \epsilon X + \frac{X^{k_c}}{k_c}
&
W_X &= - \epsilon +  X^{k_c-1} 
\nonumber \\
H &= W_{XX} =  (k_c-1) X^{k_c-2}
&
X_n &= \epsilon^{\frac{1}{k_c-1}} e^{ \frac{2 \pi i}{k_c-1} n}
\end{align}
where the $X_n$ label the $k-1$ vacua over which we sum in formula (\ref{LGformula}), and $n=0,1,\dots,k_c-2$.
At small $\epsilon$, we can then approximate e.g. the spherical one-point function by:
\begin{eqnarray}
\langle X^j \rangle_0 &=& \epsilon^{\frac{j}{k_c-1}} \sum_{n=0}^{k_c-2} e^{ \frac{2 \pi i}{k_c-1} j n}
( k_c-1)^{-1} e^{ \frac{-2 \pi i}{k_c-1} (k_c-2)  n} \epsilon^{-\frac{k_c-2}{k_c-1}} \, .
\end{eqnarray}
For a positive integer power $j$ we find that this equals 
\begin{eqnarray}
\langle X^j \rangle_0 &=& \delta_{j,k_c-2}  \, .
\end{eqnarray}
This agrees with the metric and three-point function on the sphere provided earlier. The regularization procedure
is unambiguous, and finite. Let's explore where the analogous approach for the non-compact models leads us.

\subsubsection{Near the Non-Compact TCFT}
We wish to find a similar confirmation for the correlators proposed for the non-compact topological conformal field 
theory in subsection \ref{noncompactTCFT}.
The first issue to address for the non-compact models is how to regulate. We can imagine various regulators: 
the logarithmic regulator of \cite{Girardello:1990sh}, a mass term linear
in $Y$, the addition
of an almost-normalizable operator $Y^{-1}$ term to the superpotential, or the addition of a normalizable $Y^{-2}$ term.
Let's discuss some of these regulators in detail.

\subsubsection*{The $Y$ Deformation}
The derivative of the superpotential is of order $k+1$ in $Y^{-1}$, and thus, we may  expect a $k+1$ fold degeneracy of the vacuum state.
If we add a term linear in the $Y$ variable, this expectation is borne out, and we calculate
\begin{align}
W&=  \epsilon Y + \frac{Y^{-k}}{k} \label{lineardeformation}
&
W_Y &=  \epsilon -  Y^{-k-1} 
\nonumber \\
H &= W_{YY} = (k+1) Y^{-k-2}
&
Y_n^{-1} &= \epsilon^{\frac{1}{k+1}} e^{ \frac{2 \pi i}{k+1} n}
\end{align}
where $n$ labels the vacua over which we sum in formula (\ref{LGformula}), and $n=0,1,\dots,k$.
We then find the spherical one-point function
\begin{eqnarray}
\langle Y^{-j} \rangle_0 &=& \epsilon^{\frac{j}{k+1}} \sum_{n=0}^{k} e^{ \frac{2 \pi i}{k+1} j n}
( k+1)^{-1} e^{- \frac{2 \pi i}{k+1} (k+2)  n} \epsilon^{-\frac{k+2}{k+1}} 
\nonumber \\
& =& \frac{1}{\epsilon} \delta_{j,1} + \delta_{j,k+2} \, .
\label{lindef}
\end{eqnarray}
If we concentrate on two-point functions of operators that are in the spectrum, we find agreement with the topological metric $\eta_{ij}$ proposed
in subsection \ref{noncompactTCFT},
as well as with the three-point functions. There is an important phenomenon to note here. A standard reasoning says that,
since we sum over points where $W_Y=0$ in the formula (\ref{LGformula}), we should expect that the vacuum expectation value of
$W_Y=-Y^{-k-1}$ as well as its multiples equal zero. Again, naively, this simple reasoning contradicts the result (\ref{lindef}). Let us explain then why on second thought, it does not. Consider the operator 
$-Y^{-1} W_Y=- \epsilon Y^{-1} + Y^{-k-2}$ after deformation by $\epsilon$. The vacuum expectation value 
of this operator is indeed zero, in an interesting manner. The vacuum expectation value of $Y^{-1}$ diverges and cancels the contribution
of the $Y^{-k-2}$ operator to the expectation value. The regularization and  an intricate cancellation leads to the possibility of a non-zero expectation value
for $Y^{-k-2}$. A similar remark applies to the regularization approaches that follow.

\subsubsection*{The Logarithmic Deformation}
We consider another regularization, amongst other reasons to check the robustness of our conclusions.
Let's review the approach of \cite{Girardello:1990sh} in the $Y$-variables.  The superpotential is deformed by a logarithmic term in the $Y$-variable.
The calculation goes
\begin{eqnarray}
W &=& \epsilon  \log Y + Y^{-k}/k \label{logdeformation}
\nonumber \\
W_Y &=& \epsilon Y^{-1} - Y^{-k-1}
\nonumber \\
W_{YY} &=& -\epsilon Y^{-2}+(k+1) Y^{-k-2}
\end{eqnarray}
and
\begin{eqnarray}
W_Y &=& 0
\end{eqnarray}
has solutions
\begin{eqnarray}
Y^{-1} &=& 0 \nonumber \\
Y_n^{-1} &=& \epsilon^{1/k} e^{2 \pi i n/k}  
\, .
\end{eqnarray}
 We have brought only $k$ of the vacuum solutions to a finite value and need to be careful about the contribution of the vacuum at infinity. 
We   apply the formula (\ref{LGformula}) and find for the spherical one-point function, for $j>2$
\begin{eqnarray}
\langle Y^{-j} \rangle &=& \sum_{n=0}^{k-1}
\epsilon^{j/k} e^{ 2 \pi i n (j-2) /k}
\epsilon^{-1-2/k} \frac{1}{k}
\nonumber \\
&=& \delta_{j,k+2}
\, .
\end{eqnarray}
For $j \le 2$, the contribution at infinity makes for an ambiguous result. For the one-point function of most interest, the result agrees
with the linear regularization. 

\subsubsection*{The $\Phi$-variables}
Note that from the perspective of the $\Phi$ variables, the contribution of the point at $Y^{-1}=0$
is  spurious, namely, solely due to the change of variables. 
Indeed, let's illustrate this with an explicit calculation in the $\Phi$ variable.
The superpotential becomes \cite{Cecotti:1991me}
\begin{equation}
W = \mu e^{b \Phi}  - m \Phi.
\end{equation}
The computation goes
\begin{eqnarray}
W_\Phi &=& \mu b e^{b \Phi}  - m
\nonumber \\
W_{\Phi \Phi} &=& \mu b^2 e^{b \Phi} 
\end{eqnarray}
There are only $k$ vacua solving the equation $\mu b (e^{ \frac{1}{2b} \Phi})^{k} = m$, as in the Witten index calculation \cite{Cecotti:1991me}. 
We find the topological metric for the operators $e^{ \frac{j}{b} \Phi}$
\begin{eqnarray}
\eta_{ij} &=& \frac{1}{mb } \sum_{n=0}^{k-1} e^{\frac{2\pi i n( i+j)}{k}} (\frac{m}{\mu b})^{\frac{i+j}{k}} = 2k \delta_{i+j,k} \, . \label{topphi}
\end{eqnarray}
The relative normalization of the bilinear form $\eta_{ij}$ is $2k$ times larger in the $\Phi$ variables than in the $Y$-variables, merely due to a change
of variables. The shift in the indices is due to the difference in measure (or Hessian).
  Indeed, we see that the topological metric is now automatically
evaluated in the state with conformal dimension and R-charge equal to $(1/2k,1/k)$ (from the shift in the indices
$i,j$ by one in the metric (\ref{topphi}) compared to (\ref{topY})). (See the discussion in subsection \ref{noncompactTCFT}.)
Other deformation proposals, based on the almost normalizable operator $Y^{-1}$ or the normalizable operator $Y^{-2}$ confirm the robust nature of the proposed
topological conformal field theory correlators. We therefore posit these correlators as our starting point and 
move on to uncover interesting properties of these topological theories.

\section{The Topological Massive  Model}
\label{TQFT}
Superconformal field theories can be deformed to massive supersymmetric quantum field theories by adding an operator
to the Lagrangian. When the deformations are consistent with $N=(2,2)$ supersymmetry, the resulting theory can
be topologically twisted to produce a topological quantum field theory. In this section, we study relevant deformations
of the $N=2$ Liouville conformal field theory that preserve supersymmetry, and twist them into topological massive
quantum field theories. The perturbed correlators are highly constrained, and governed by a free energy determined
by WDVV equations. The deformation space is a Fr\"obenius manifold \cite{Dubrovin:1994hc}. For compact models, the deformed operators
with given R-charge, as well as the structure constants as a function of the deformation parameters were determined in 
\cite{Dijkgraaf:1990dj}. We do refer to this reference for details on the solution of the compact model, which we only summarize below.
The solution of the non-compact model is more intricate still, and we will describe it in detail.
\subsection{The Compact Topological Field Theory}
\label{compactTQFT}
For the $N=2$ minimal model and its relevant deformation, a solution for the topological theory was elegantly presented in
\cite{Dijkgraaf:1990dj}.
The deformed structure constants $c_{ijl}$ of the ring of the compact theory are defined in terms of the correlation functions of operators $\phi_i$ at criticality and with fixed R-charge,
as follows:
\begin{equation}
c_{ijl}(t) = \langle \phi_i \phi_j \phi_l \exp (\sum_n t_n \int \phi_n) \rangle_0 \, ,
\end{equation}
where the correlators are evaluated in the topological conformal theory and where the deformation of the action is taken into account through the
exponentiation of the integrated operators \cite{Dijkgraaf:1990dj}.
The structure constants satisfy a number of consistency conditions that imply that they can be derived from a generating functional $F$ which is the
free energy at tree level \cite{Dijkgraaf:1990dj}.
Two paths to the calculation of the structure constants are laid out in \cite{Dijkgraaf:1990dj}. The first is to define a deformed superpotential
\begin{equation}
W(X,t) = \frac{1}{k_c} X^{k_c} - \sum_{i=0}^{k_c-2} g_i(t_j) X^i \, ,
\end{equation}
where the $g_i$ are equal to the deformation parameters $t_i$ to first order in $t_i$. The chiral primaries $\phi_i$ are equal to the 
original chiral primaries $X^i$ at zero deformation, and  have a fixed R-charge $q_i=i/k$, as well
as fixed leading order behavior $X^i$. They are proposed to become parameter dependent as follows 
\begin{equation}
\phi_i(X;t) = - \partial_i W(X,t) \, ,
\end{equation}
where $\partial_i$ is a partial derivative with respect to $t_i$.
The chiral ring coupling constants ${c_{ij}}^l(t)$ are then computed with respect to the basis $\phi_i$, i.e. they satisfy $\phi_i \phi_j={c_{ij}}^l \phi_l$ 
for all $t_i$.

On the ring $R_c=\mathbb{C}[X]/\langle W_X \rangle$ of polynomials modulo the derivative of the superpotential,
a non-degenerate symmetric bilinear form is introduced which equals a contour integral of a ratio of polynomials
\begin{equation}
(\chi_1,\chi_2) = \sum Res ( \frac{\chi_1 \chi_2}{W_X}) = \oint \frac{dX}{2 \pi i} \frac{\chi_1 \chi_2}{W_X} \, .
\label{eq:bilinear}
\end{equation}
The contour is taken to circle  all zeroes of $W_X$, and can be taken  large (compared to the zeros).
The bilinear form is independent of the chosen representatives $\chi_i$ in the quotient ring $R_c$. Given the leading behavior of the operators $\phi_i$, and expanding
near infinity, we can prove \cite{Dijkgraaf:1990dj} that the 
metric equals
\begin{equation}
(\phi_i,\phi_j) = {c_{ij}}^l \sum Res (\phi_l/W_X) = c_{ij0} = \eta_{ij} \, .
\end{equation}
The structure constants are then given in terms of the $\phi_i$ by the formula
\begin{equation}
c_{ijl}(t) = \sum Res \frac{ \phi_i \phi_j \phi_l}{W_X} \, .
\end{equation}
Crucially, by exploiting a parafermionic selection rule that is inherited from $SU(2)_{k_c}$ fusion rules, a recursion relation for the operators $\phi_i$ of fixed R-charge
is proven in \cite{Dijkgraaf:1990dj}, which permits
a closed form solution. It is given by the determinant formula  
\begin{equation}
\phi_i(X,t) = (-1)^i \det \left( \begin{array}{ccccc}
-X & 1 & 0 & \dots & 0 \\
t_{k_c-2} & -X & 1 & \dots & \dots \\
t_{k_c-3} & t_{k_c-2} & \dots & \dots & 0 \\
\dots & \dots & \dots & \dots & 1 \\
t_{k_c-i} & \dots & t_{k_c-3} & t_{k_c-2} & -X 
\end{array} \right) \, , \label{compactphis}
\end{equation}
for $i=0,1, \dots , k_c-2$.
One also obtains
\begin{equation}
W_X (X,t) = \phi_{k_c-1} (X,t) \, ,
\end{equation}
where the right hand side extends the definition (\ref{compactphis}) to $i=k_c-1$.
One can then integrate up to find  $W(X,t)$, using also the differential equations $\phi_i=-\partial_i W$. 

A second manner in which the solution is obtained is by defining the auxiliary (Lax) root $L$
of the superpotential
\begin{equation}
\frac{L^{k_c}}{k_c} = W \, ,
\end{equation}
and to think of it as a formal power series
\begin{equation}
L(X) = X (1+ \sum_{j=2} b_j X^{-j})
\end{equation}
near large $X$. The fields $\phi_i$ are then proposed to be
\begin{equation}
\phi_i(X) = [L^i \partial_X L]_{\ge 0}
\end{equation}
where the index on square brackets indicates which terms in the formal power series are to be taken. 
The residues of the root $L$ are the parameters $t_i$:
\begin{equation}
t_{k_c-2-i} = -\frac{1}{i+1}  Res (L^{i+1}) \, .
\end{equation}
Since we expanded at large $X$ (e.g. by imagining a large contour, i.e. looking at the residues from afar), we can evaluate the residue as the coefficient of the $1/X$ term.
The consistency of the first and the second method can be demonstrated using the Lagrange inversion formula for formal power series expansions,
reviewed in appendix \ref{formalpowerseries}.

Furthermore, the structure constants can then be computed to be \cite{Dijkgraaf:1990dj}
\begin{equation}
c_{ijl} = \frac{1}{(l+1)(k_c+l+1)} \partial_i \partial_j Res(L^{k_c+l+1}) \, ,
\end{equation}
which implies that the first derivatives of the free energy integral $F=\langle \exp(\sum_i t_i \phi_i) \rangle_0$  equal 
\begin{equation}
\partial_i F = \frac{1}{(i+1)(k_c+i+1)} Res (L^{k_c+i+1}) \, .
\end{equation}
The proofs of this laundry list of formulas are non-trivial and interesting, and they are given in \cite{Dijkgraaf:1990dj}, when combined with the Lagrange inversion formula. 
We will provide  proofs for the non-compact
models below, in great detail.
Our goal in reviewing the compact formulas is to have a simpler and useful point
of comparison. Explicit examples of generators of correlation functions $F$ are integrated  in appendix \ref{compactexamples} for levels $k_c=2,3,4$ and $5$.

\subsection{The Non-Compact Topological Field Theory}
In this section, we study relevant deformations of the topological non-compact conformal field theory that we introduced in section \ref{TCFT}. We limit ourselves to strictly
normalizable deformations, proportional to an operator in the ring $R$ generated by the set of operators $\{ Y^{-2}, Y^{-3}, \dots, Y^{-k} \}$. The selection rules on fusion are
less strong in the non-compact model, and therefore we only follow the second path reviewed in subsection \ref{compactTQFT}.
We firstly propose a closed form solution, and then verify that it satisfies all  necessary conditions. Our first goal is
to find deformed operators $\phi_i$ of fixed R-charge $i/k$ and with leading behavior $Y^{-i}$. The  deformation parameters of the model will be denoted $t_{i}$.
We also make use of the alternative notation $t_{k-i}=-s_{i=1,\dots,k-2}$ and $t_k=1-s_0$,
where the label on $s_i$ equals its R-charge (times the level $k$). Note that there is a marginal operator in the ring $R$, and there is therefore a parameter $s_0$ with
zero R-charge. Moreover, we will take the parameter $s_0$ to have value one in the undeformed theory while the other parameters are zero. We impose the constraint
that the parameter $s_0$ is always non-zero.

Secondly, we wish to determine the superpotential (or at least, all of its derivatives),
and thirdly, we compute the structures constants $c_{ijl}$ which are again generated by deformation
\begin{equation}
c_{ijl}(t) = \langle \phi_i \phi_j \phi_l \exp (\sum_{n } t_n \int \phi_n) \rangle_0 \, ,
\end{equation}
where the operators on the right hand side are those at the conformal point.
 Finally, we calculate the free energy $F$ from which the structure constants and all correlators can be derived.
\subsubsection*{The Operators}
The bilinear form on the deformed ring, in analogy to equation \eqref{eq:bilinear}, is given by
\begin{eqnarray}
( \chi_1 , \chi_2 ) &=& \sum Res_Y ( \frac{ \chi_1 \chi_2}{W_Y}) \, ,
\end{eqnarray}
where the sum is over all the zeros of the derivative of the superpotential $W_Y$. Since the zeros of $W_Y$ are near $Y^{-1}=0$, we
rewrite the formula in terms of the inverse $Y^{-1}$
\begin{eqnarray}
( \chi_1 , \chi_2 ) &=& -\sum Res_{Y^{-1}} ( \frac{Y^2 \chi_1 \chi_2}{W_Y}) \, .
\end{eqnarray}
The symmetry and bi-linearity of the form $(\cdot,\cdot)$ are obvious. The independence of the choice of representative follows from the fact
that the ring elements have at least a $Y^{-2}$ prefactor, and only positive powers of $Y^{-1}$. Non-degeneracy will follow from our explicit choice
of generators of the ring, and the calculation of the non-degenerate metric $\eta$.

The deformed superpotential for the massive models can be generically parameterized as
\begin{eqnarray}
W 
  &=& \frac{Y^{-k}}{k} (1+\sum_{i=0}^{k-2} h_{i} Y^i) \, .
\end{eqnarray}
The functions $h_{i}$ when expanded to first order in all   $t_j$ only have a non-zero coefficient in front of  $t_{k-i}$. The leading coefficient $1+h_0$ must remain non-zero in order to keep the asymptotics of the model intact.
Following the lead of the compact model, we define the root $L$ of the superpotential
\begin{equation}
L = (kW)^{\frac{1}{k}} \, ,
\end{equation}
which we think of as a formal power series in $Y$,
and propose the equation
\begin{equation}
\phi_i = {[} -L^{i-2} \partial_Y L {]}_{\le -2} \label{defphi}
\end{equation}
for the operators $\phi_i$ of definite charge $q_i=i/k$ in the deformed chiral ring.
Moreover, we define the deformation parameters as
\begin{equation}
t_{k-i} - \delta_{i,0} \equiv -s_i = -\frac{1}{i+1} Res_{Y^{-1}} (Y^2 L^{i+1}) \label{ss}
\, .
\end{equation}

We must prove the following properties. The operators $\phi_i$ have the appropriate charge. 
The metric $\eta_{ij}$ is undeformed. 
The structure constants $c_{ijl}$ associated to the operators $\phi_i$ obey that they are symmetric, 
and they can be integrated up to a free energy functional $F$.
We will prove these properties one by one.

Firstly, the charge of the root $L$ is $1/k$. The charge of the operator $Y$ is $-1/k$. Thus, the field $\phi$ in equation (\ref{defphi}) has charge
$i/k$ as desired. The leading behavior of $\phi_i$ is dictated by the leading behavior of $L$, and is indeed $Y^{-i}$, with lower
powers of $Y^{-1}$ trailing. The lower powers are all in the strict chiral ring $R$ by the design of equation (\ref{defphi}).

The calculation of the metric in the deformed chiral ring generated by the operators $\phi_i$ for $i=2,3,\dots,\le k$ reads
\begin{eqnarray}
(\phi_i,\phi_j) &=&   Res_{Y^{-1}} (Y^2  {[} L^{i-2} \partial_Y L {]}_{\le -2}  \phi_j / W_Y)
\nonumber \\
&=&  Res_{Y^{-1}} (Y^2 L^{i-k-1}\phi_j)
\nonumber \\
&=&  Res_{Y^{-1}} (Y^2 L^{i-k-1}{[} (-)L^{j-2} \partial_Y L {]}_{\le -2})
\nonumber \\
&=&  Res_{Y^{-1}} (-Y^2  L^{i+j-k-3} \partial_Y L) 
\nonumber \\
&=& \delta_{i+j,k+2}\, ,
\end{eqnarray}
where the first step is true due to the fact that we can drop the constraint on the terms we consider in the bracket since  
\begin{equation}
Y^2 {[} L^{i-2} \partial_Y L {]}_{> -2}  \phi_j / W_Y=Y^{k-j+3}\times(\text{sum of non-negative powers of }Y)
\end{equation}
does not contain a linear term in $Y$.
We have used that $L^{i-2} \partial_Y L=\frac{1}{i-1}\partial_Y L^{i-1}$ does not have a $Y^{-1}$ term for $i\neq1$.
The second step uses the definition (\ref{defphi}) of the operator $\phi_j$. 
The third step is justified by a similar argument as the first step.
The last step is made while observing that only when $i+j-k-3=-1$ the argument of the residue is not a total derivative,
and that in that case, the residue is one. Thus, the metric is invariant under deformation, and non-degenerate.

We also require  that after deformation, the equation $\partial_i W = - \phi_i$ is still valid.  By deriving the  equation for $W$ in terms of the root $L$, 
and plugging the resulting equation for $\phi_i=-L^{k-1} \partial_i L$ (where the derivative is with respect to $t_i$) into the equation for the metric $\eta$:
\begin{eqnarray}
\eta_{ij} &=& Res_{Y^{-1}}(Y^2 L^{i-1-k} \phi_j) = -\frac{1}{i-1} Res_{Y^{-1}} \partial_i (Y^2 L^{i-1}) \, ,
\end{eqnarray}
we find after integration that equation (\ref{ss}) holds (where $t_i$ are conveniently compared to the compact models, while the alternative parameterization
in terms of $s_i$ is handy for keeping track of R-charges). 

\subsubsection*{The Explicit Operators}
In principle, these equations are sufficient description of the solution of the model, but it is interesting to solve the model more explicitly. In particular,
we are confronted with a considerably more intricate combinatorics problem than was the case in the compact model of subsection \ref{compactTQFT}. We delve into
this combinatorial quagmire, and compute the operators $\phi_i$ as explicit expressions in the  parameters $s_i$.
Our starting points are equations (\ref{defphi}) and (\ref{ss}). 

Firstly, notice that from the superpotential and the definition of the root $L$, we can write
\begin{equation}
YL = \sum_{n=0}^\infty b_n Y^n \, ,
\end{equation}
where the coefficient $b_0$ satisfies $b_0\neq 0$. The equation (\ref{ss}) for the parameters $s_i$  then becomes
\begin{eqnarray}
s_{j} &=& \frac{1}{j+1} [ (YL)^{j+1}]_{j} 
\nonumber \\
&=&   \frac{b_0^{j+1}}{j+1} [1+\sum_{l=1}^{j+1} \binom{j+1}{l} (\sum_{n=1}^\infty \frac{b_n}{b_0} Y^n)^l ]_{j}
\nonumber \\
&=&    \frac{b_0^{j+1}}{j+1}\sum_{l=1}^{j} \binom{j+1}{l}\frac{l!}{j!}B_{j,l}(1!\frac{b_1}{b_0},2!\frac{b_2}{b_0},\cdots)
\nonumber \\
&=&   \sum_{l=1}^{j} \frac{b_0^{j+1}}{(j+1-l)!}B_{j,l}(1!\frac{b_1}{b_0},2!\frac{b_2}{b_0},\cdots)
\end{eqnarray}
for $j\neq 0$, where the polynomials $B_{j,l}$ are partial Bell Polynomials \cite{Bell,Comtet} that arise from expanding powers of multinomials. 
Remark that for the special case $j=0$, one finds $s_0=b_0$.
We then get for $j\neq 0$,
\begin{equation}
j!\frac{s_j}{s_0^{j+1}}=\sum_{l=1}^{j} \binom{j}{l-1}(l-1)!B_{j,l}(1!\frac{b_1}{b_0},2!\frac{b_2}{b_0},\cdots)\, .
\label{bintermsofs}
\end{equation}
The combinatorial heart of the calculation occurs at the following step, which, given equation (\ref{bintermsofs}),  relates the partial Bell polynomials
in the parameters $s$ with the partial Bell polynomials of the coefficients $b$ of the operator $YL$. This step is given by Theorem 15 of \cite{Birmajer}, which states that for all integers $1 \leq p<i$,
we now have
\begin{equation}
\sum_{l=1}^{p}\binom{i-p-1}{l-1}(l-1)!B_{p,l}(1!\frac{s_1}{s_0^2},2!\frac{s_2}{s_0^3},\cdots)=\sum_{l=1}^{p}\binom{i-1}{l-1}(l-1)!B_{p,l}(1!\frac{b_1}{b_0},2!\frac{b_2}{b_0},\cdots)\, .\label{eq:Bell}
\end{equation}
Similar to the above computation for the parameters $s_j$, the operators (\ref{defphi}) become
\begin{eqnarray}
\phi_{i+1} &=& - \frac{1}{i}  {[} \partial_Y ( Y^{-i} s_0^i (1+ \sum_{n=1}^\infty \frac{b_n}{b_0} Y^n)^i ) {]}_{\le -2}
\nonumber \\
&=& s_0^i[Y^{-i-1}+\sum_{p=1}^{i-1} f^{(i)}_p Y^{-i-1+p}]\, ,
\end{eqnarray}
where 
\begin{eqnarray}
f^{(i)}_p &=& \frac{i-p}{i}[(1+ \sum_{n=1}^\infty \frac{b_n}{b_0} Y^n)^i]_p
\nonumber \\
&=& \frac{i-p}{i}\sum_{l=1}^{p}\binom{i}{l}\frac{l!}{p!}B_{p,l}(1!\frac{b_1}{b_0},2!\frac{b_2}{b_0},\cdots)
\nonumber \\
&=& \frac{i-p}{p!}\sum_{l=1}^{p}\binom{i-1}{l-1}(l-1)!B_{p,l}(1!\frac{b_1}{b_0},2!\frac{b_2}{b_0},\cdots)\, .
\end{eqnarray}
Multiplying by $\frac{i-p}{p!}$ on both sides of \eqref{eq:Bell} one obtains
\begin{eqnarray}
f^{(i)}_p &=& \sum_{l=1}^{p} \binom{i-p}{l}\frac{l!}{p!}B_{p,l}(1!\frac{s_1}{s_0^2},2!\frac{s_2}{s_0^3},\cdots)\, .
\end{eqnarray}
Now we use the definition of the Bell polynomials \cite{Comtet} and get
\begin{equation}
\frac{l!}{p!}B_{p,l}(1!\frac{s_1}{s_0^2},2!\frac{s_2}{s_0^3},\cdots)=\sum \frac{l!}{r_1!\cdots r_{p-l+1}!}(\frac{s_1}{s_0^2})^{r_1}\cdots (\frac{s_{p-l+1}}{s_0^{p-l+2}})^{r_{p-l+1}}\, ,
\label{defBell}
\end{equation}
where the sum is taken over all non-negative integers $r_1,r_2,\cdots,r_{p-l+1}$ such that $\sum_{n=1}^{p-l+1}r_n=l$ and $\sum_{n=1}^{p-l+1}nr_n=p$. The right hand side of equation (\ref{defBell})
can be rewritten
as
\begin{equation}
s_0^{-p-l}\times\sum_{j_1+\cdots+j_l=p,j_n\geqslant 1}s_{j_1}\cdots s_{j_l}\, .
\end{equation}
Therefore, we find the coefficients $f^{(i)}_p$ of the operators
\begin{eqnarray}
f^{(i)}_p &=& s_0^{-i}\sum_{j_1+\cdots+j_{i-p}=p,j_n\geqslant 0} s_{j_1}\cdots s_{j_{i-p}}\, .
\end{eqnarray}
In summary, the explicit expression for the operators $\phi_{i+1}$ in term of the parameters $s_j$ is 
\begin{eqnarray}
\phi_{i+1} &=& \sum_{p=0}^{i-1}\sum_{j_1+\cdots+j_{i-p}=p} s_{j_1} \cdots s_{j_{i-p}} Y^{-i-1+p}\, .
\label{explicitoperators}
\end{eqnarray}

There is also an interesting recursion relation among the operators $\phi_i$:
\begin{equation}
\phi_{i+2}=s_0Y^{-1}\phi_{i+1}+\sum_{n=1}^i\left[-\sum_{l=1}^n(-s_0)^{-l}\sum_{j_1+\cdots+j_l=n,j_m\geqslant 1}s_{j_1}\cdots s_{j_l}\right]\phi_{i+2-n}\, .
\label{eq:recursion}
\end{equation}
We prove this recursion relation by comparing the coefficient of $Y^{-i-2+p}$ for all $p$ from $0$ to $i$ on both sides of  equation (\ref{eq:recursion}). For $p=0$ it is trivial that the two coefficients match. For $p>0$, we proceed as follows. We first notice that the coefficient in front of the operator $\phi_{i+2-n}$ in the recursion relation (\ref{eq:recursion})
can be written in terms of the partial Bell polynomials as
\begin{equation}
s_0^n\sum_{l=1}^n(-)^{l-1}\frac{l!}{n!} B_{n,l}(1!\frac{s_1}{s_0^2},2!\frac{s_2}{s_0^3},\cdots)\, .
\end{equation}
The coefficient of $Y^{-i-2+p}$ on the left hand side of  equation (\ref{eq:recursion}) is then
\begin{equation}
s_0^{i+1} \sum_{l=1}^p \binom{i+1-p}{l}\frac{l!}{p!} B_{p,l}(1!\frac{s_1}{s_0^2},2!\frac{s_2}{s_0^3},\cdots)\, ,
\label{eq:coefficientl}
\end{equation}
while the coefficient on the right hand side is the sum of
\begin{equation}
s_0^{i+1} \sum_{l=1}^p \binom{i-p}{l}\frac{l!}{p!} B_{p,l}(1!\frac{s_1}{s_0^2},2!\frac{s_2}{s_0^3},\cdots)
\label{eq:coefficientr1}
\end{equation}
and
\begin{equation}
s_0^{i+1}\sum_{n=1}^p\sum_{l_1=0}^{p-n}\sum_{l_2=1}^n \binom{i+1-p}{l_1}\frac{(-)^{l_2-1}l_1!l_2!}{(p-n)!n!}B_{p-n,l_1}(1!\frac{s_1}{s_0^2},2!\frac{s_2}{s_0^3},\cdots)B_{n,l_2}(1!\frac{s_1}{s_0^2},2!\frac{s_2}{s_0^3},\cdots)\, .
\label{eq:coefficientr2}
\end{equation}
Using the formula $B_{p,l}=0$ for $p<l$, we do not change the value of expression \eqref{eq:coefficientr2} if we change the range of the two sums over $l_1$ and $l_2$  into $\sum_{l_1=0}^{p-1}$ and $\sum_{l_2=1}^{p}$. We can then do the sum over $n$ by applying  equation (1.4) of \cite{Cvijovic} for the partial Bell polynomials and obtain
\begin{equation}
s_0^{i+1}\sum_{l_1=0}^{p-1}\sum_{l_2=1}^p \binom{i+1-p}{l_1} \frac{(-)^{l_2-1}(l_1+l_2)!}{p!} B_{p,l_1+l_2}(1!\frac{s_1}{s_0^2},2!\frac{s_2}{s_0^3},\cdots)\, 
\end{equation}
for this part of the coefficient of $Y^{-i-2+p}$ on the right hand side of (\ref{eq:recursion}).
Furthermore, denote $l=l_1+l_2$ such that the expression \eqref{eq:coefficientr2} becomes
\begin{equation}
s_0^{i+1}\sum_{l=1}^p\left[(-)^{l-1}\sum_{l_1=0}^{l-1}(-)^{l_1}\binom{i+1-p}{l_1}\right] B_{p,l}(1!\frac{s_1}{s_0^2},2!\frac{s_2}{s_0^3},\cdots)\, .
\end{equation}
Since we have the binomial identity
\begin{equation}
\sum_{l_1=0}^{l-1}(-)^{l_1}\binom{i+1-p}{l_1}=(-)^{l-1}\binom{i-p}{l-1}\, ,
\end{equation}
we  find that expression \eqref{eq:coefficientr2} is equal to
\begin{equation}
s_0^{i+1} \sum_{l=1}^p \binom{i-p}{l-1}\frac{l!}{p!} B_{p,l}(1!\frac{s_1}{s_0^2},2!\frac{s_2}{s_0^3},\cdots)\, .
\end{equation}
Finally, adding the contribution \eqref{eq:coefficientr1} to the coefficient of the right hand side,  we find that the total coefficient of $Y^{-i-2+p}$ in the recursion relation
\eqref{eq:recursion} is indeed equal to the left hand side coefficient \eqref{eq:coefficientl}, which proves the recursion relation.
We believe this relation is  interesting by itself, since a comparison to the compact case suggests that they may follow from non-compact fusion rules.

The combinatorial theorems \cite{Birmajer,Cvijovic} that are central to this subsection are surprisingly intricate and recent, but they are proven with elementary analytic means. 
It would be good to have a better conceptual understanding of these combinatorial theorems,
and a clearer picture of what is counted e.g. in graph theoretical terms. Instead, we concentrate on the further description of our solution to the non-compact model.


\subsubsection*{The Superpotential}
Similarly as in the compact case, the $Y$-derivative of the superpotential is given by the expression for the $\phi_i$,
where we equate the index to $i=k+1$. We thus have the formula
\begin{equation}
W_Y = -\phi_{k+1} \, .
\end{equation}
We can then integrate the equations $\partial_i W = -\phi_i$ as well as $W_Y=-\phi_{k+1}$ 
to obtain an expression for the superpotential (up to an arbitrary constant). 

\subsubsection*{The Structure Constants and the Free Energy}

The equations (\ref{explicitoperators}) also provide solutions for the structure constants ${c_{ij}}^l(s)$ as a function of the  parameters $s_i$,
through the equations
\begin{eqnarray}
\phi_i \phi_j &=& {c_{ij}}^l (s) \phi_l \, .
\end{eqnarray}
We can prove that the definition of the  parameters $s_i$ is consistent with desired properties of
the structure constants $c_{ijl}$, namely, their symmetry and integrability, and we will be able to
generically integrate for the first derivatives of the generating functional $F$. Derivatives $\partial_i$ will again
be with respect to the variables $t_i$.
As a preliminary, we derive the useful equation
\begin{equation}
\partial_j \phi_i = \partial_j [ -\frac{1}{i-1} \partial_Y L^{i-1}]_{\le -2}
\nonumber \\
=- \frac{1}{i-1} \partial_Y ( [\partial_j L^{i-1}]_{\le -1})
\nonumber \\
= - \partial_Y ([ L^{i-2} \partial_j L]_{\le -1}) \, .
\end{equation}
Next, we demonstrate that the three-point function, which is manifestly symmetric, is a double derivative with respect to the variables $t_i$:
\begin{eqnarray}
c_{ijl} &=& -Res_{Y^{-1}} (\frac{Y^2}{W_Y} \phi_i \phi_j \phi_l)
= Res_{Y^{-1}} (\frac{Y^2}{W_Y} \phi_i \phi_j [L^{l-2} \partial_Y L]_{\le -2} )
\nonumber \\
&=& Res_{Y^{-1}} (Y^2 \phi_i \phi_j L^{-k+l-1})- Res_{Y^{-1}} (\frac{Y^2}{W_Y} \phi_i \phi_j [L^{l-2} \partial_Y L]_{\ge -1} )
\nonumber \\
&=& \frac{1}{l-1} ( -Res_{Y^{-1}} (Y^2 \phi_i  \partial_j L^{l-1} )- Res_{Y^{-1}} (Y^2 [\frac{\phi_i \phi_j}{W_Y}]_{\le -1} \partial_Y (L^{l-1}) ))
\nonumber \\
&=& \frac{1}{l-1} ( -Res_{Y^{-1}} (Y^2 \phi_i  \partial_j L^{l-1} )+ Res_{Y^{-1}} (Y^2 \partial_Y [\frac{\phi_i \phi_j}{W_Y}]_{\le -1} L^{l-1} ))
\nonumber \\
&=& \frac{1}{l-1} ( -Res_{Y^{-1}} (Y^2 \phi_i  \partial_j L^{l-1} )- Res_{Y^{-1}} (Y^2 \partial_Y [L^{i-k-1} \phi_j]_{\le -1} L^{l-1} ))
\nonumber \\
&=& -\frac{1}{l-1}  Res_{Y^{-1}} (Y^2  \partial_j (\phi_i  L^{l-1}) )
= \frac{1}{(l-1)(l+k-1)} \partial_i \partial_j Res_{Y^{-1}} (Y^2 L^{k+l-1}) \, ,
\end{eqnarray}
where we again restricted to $2 \le i,j,l \le k$. The justification for the steps in the proof are of the same type as those we detailed for the calculation
of the topological metric $\eta_{ij}$. We then find that the first derivatives of the generator
of correlation functions $F$ is
\begin{eqnarray}
\partial_l F &=& \frac{Res_{Y^{-1}}(Y^2 L^{k+l-1})}{(l-1)(l+k-1)} \, . \label{Fderivatives}
\end{eqnarray}
Crucially, the  equation (\ref{Fderivatives}) is integrable. The proof of the integrability proceeds as follows:
\begin{eqnarray}
\partial_j \partial_l F &=& -\frac{Res_{Y^{-1}}(Y^2 L^{l-1} \phi_j)}{l-1}
\nonumber \\
&=&  \frac{Res_{Y^{-1}}(Y^2 L^{l-1} [\partial_Y L^{j-1} ]_{\le -2})}{(l-1)(j-1)}
\nonumber \\
&=&  \frac{Res_{Y^{-1}}(Y^2 [L^{l-1}]_{\ge 1} \partial_Y L^{j-1} )}{(l-1)(j-1)}
\nonumber \\
&=& -\frac{Res_{Y^{-1}}(Y^2 [\partial_Y L^{l-1}]_{\ge 0}  L^{j-1} )}{(l-1)(j-1)}
\nonumber \\
&=& \frac{Res_{Y^{-1}}(Y^2 [\partial_Y L^{l-1}]_{\le -2}  L^{j-1} )}{(l-1)(j-1)}
\nonumber \\
&=& \partial_l \partial_j F \, .
\end{eqnarray}
We have used the fact that the residue of a total derivative is zero twice in the previous to last step, once for $L^{l-1}$ and once for $L^{l+j-2}$.

Finally, in appendix \ref{noncompactexamples} we provide explicitly integrated values of the generating function $F$ for the non-compact models at levels $2,3,4$ and $5$ .

\subsubsection*{Final Remarks}
The solution of the massive non-compact topological quantum field theory through the determination of its chiral ring reveals novel features. It generalizes the  integrable structure that existed
in the compact case in a beautiful and non-trivial manner. We needed to enlarge our combinatorial toolkit in order to find the explicit solution, in particular to invert a power series expansion
in terms of partial Bell polynomials. It will  be interesting to also manifestly link our solution to an integrable hierarchy.

\section{The Conformal Metric}
\label{conformalmetric}
In this section we explore the extension of our analysis of the topological non-compact theories to the topological-anti-topological sector \cite{Cecotti:1991me}. In particular, we calculate the overlap
between states generated by chiral, and states generated by anti-chiral operators. 

The overlap between chiral and anti-chiral states defines a metric $g_{mn}$ which can be computed through localization
at the conformal point. This fact was understood long time ago \cite{Cecotti:1991me}, and recently carefully confirmed using contemporary techniques \cite{Ishtiaque:2017trm,Chen:2017fvl}.
The final formula for the metric $g_{mn}$ at the conformal point is
\begin{eqnarray}
g_{m\bar{n}}(0) &=& \frac{1}{\pi} \int dY d\bar{Y} Y^{-m} \bar{Y}^{-\bar{n}} \exp(W(Y)-\bar{W}(\bar{Y}))
\nonumber \\
&=& \frac{1}{\pi} \int dX d\bar{X} X^{m-2} \bar{X}^{n-2} \exp(W(X^{-1})-\bar{W}(\bar{X}^{-1}))
\nonumber \\
&=& \delta_{m,n}k^{\frac{2m-2}{k}-1}\frac{\Gamma(\frac{m-1}{k})}{\Gamma(1-\frac{m-1}{k})} \, ,
\label{eq:conformalmetric}
\end{eqnarray}
where we have used the result of the chiral-anti-chiral metric computation for a compact minimal model given in \cite{Cecotti:1991me,Ishtiaque:2017trm,Chen:2017fvl}, 
and constrained ourselves to the operators in the spectrum only, namely to the quantum numbers $2\leqslant m,n\leqslant k$. The change of variables
$Y=X^{-1}$ reduces the problem to the calculation  for $N=2$ minimal models, in contrast to what we saw for the topological chiral ring. The normalization of the metric
agrees with our normalization for the topological metric $\eta$, and the reality constraint to be discussed in section \ref{ttstar}.

It is interesting to compare this calculation to the calculation in the variables $\Phi$. It is crucial to note that even for constant field configurations,
on the sphere, there can be a factor in the path integral that arises from a linear dilaton term in the action.
The linear dilaton contribution in the radial direction $\rho$ is 
\begin{eqnarray}
S_{l.d.} &=& \frac{1}{4 \pi} \int d^2 \sigma \frac{Q}{2} \rho \sqrt{\hat{g}} {\cal R}^{(2)}
\end{eqnarray}
where $\hat{g}$ and ${\cal R}^{(2)}$ are the determinant of the sphere metric and the world sheet Ricci scalar.
The coefficient $Q$ is related to the central charge
by the equation
\begin{equation}
Q = \sqrt{\frac{2}{k}}=\frac{1}{b} \, .
\end{equation}
In the localization computation, only the zero modes of fields contribute. Therefore, we can take $\rho$ out from the integral and use the Gauss-Bonnet theorem to find
\begin{equation}
S_{l.d.} = \frac{\rho}{b} \, .
\end{equation}
The factor $e^{-S_{l.d.}}$ thus becomes part of the measure in the euclidean path integral.
In the $\Phi$ variables then, the metric is
\begin{eqnarray}
g_{m\bar{n}}(0) &=& \frac{1}{\pi} \int_{0\leqslant\theta\leqslant 4\pi b} d \phi d \bar{\phi} e^{-\frac{1}{b} \rho} e^{ \frac{m}{2b} \phi}e^{ \frac{\bar n}{2b} \bar{\phi}} \exp(W-\bar{W})
\nonumber \\
&=& \frac{1}{\pi} \int_{0\leqslant\theta\leqslant 4\pi b} d \phi e^{- \frac{1}{2b} \phi} d \bar{\phi} e^{- \frac{1}{2b} \bar{\phi}}
e^{ \frac{m}{2b} \phi}e^{ \frac{\bar n}{2b} \bar{\phi}} \exp(W-\bar{W}) \, ,
\end{eqnarray}
where $\rho$ and $\theta$ are the real and imaginary zero modes of the bosonic component $\phi$ of the superfield $\Phi$.
The range of $\theta$ is determined by our choice of Liouville theory, as discussed  in section \ref{TCFT}.
By a change of variables $Y=e^{-\frac{1}{2b}\phi}$ and setting $\mu=\frac{1}{k}$, the metric equals the result in \eqref{eq:conformalmetric} multiplied by $(2b)^2=2k$.
The change of normalization (also visible in the topological metric $\eta$ in section \ref{TCFT}) is due to the change of variables from $Y$ to $\Phi$, and can be absorbed
in a redefinition of the path integral measure if so desired.

We can perform a consistency check on the localization result for the metric. The structure constants ${c_{ij}}^l(0)$ at the conformal point
were computed in \cite{Girardello:1990sh} by first determining the wave-functions of the Ramond-Ramond sector ground states in a quantum mechanics approach to the 
supersymmetric quantum field theory, and then calculating the overlap of an operator sandwiched between these wave-functions. The final result for the structure
constants was
\begin{equation}
{C^{\epsilon}_{q_1 q_2}}^{q_3} = \left( \frac{\Gamma(\epsilon)}{\Gamma(1-\epsilon)}\frac{\Gamma(1-q_1-\epsilon)\Gamma(1-q_2-\epsilon)\Gamma(q_3+\epsilon)}{\Gamma(q_1+\epsilon)\Gamma(q_2+\epsilon)\Gamma(1-q_3-\epsilon)} \right)^{\frac{1}{2}} \, ,
\label{overlap}
\end{equation}
where $q_i$ denote the R-charges of the operators and  $\epsilon$ is a regulator. In \cite{Girardello:1990sh}, the regulator $\epsilon$ measures the difference 
in R-charge between a reference state and the (almost normalizable) state with R-charge $1/k$.
The choice of normalization of operators in \cite{Girardello:1990sh} was to take the metric $g_{m \bar{n}}(0)=1$, which implies that we must renormalize our trivial structure constants ${c_{i_1 i_2}}^{i_3}(0)$ in order to compare.
Moreover, we also wish to take as our evaluation state a state close to the almost normalizable state with R-charge $1/k$. Therefore, we normalize our metric
by the norm squared of the state with label $m=1+\delta$, and we consider our operator charges (acting on this state) to be $q_i=(m_i-1)/k$. See also the discussion
of this point in section \ref{TCFT}.
Under these circumstances, we find, after adapting our normalization and using (\ref{eq:conformalmetric})
\begin{equation}
{C^{\delta}_{q_1 q_2}}^{q_3} = k^{-\delta/k} \left( \frac{\Gamma(\delta/k)}{\Gamma(1-\delta/k)} \right)^{1/2} \left( \frac{\Gamma(1-q_1)\Gamma(1-q_2)\Gamma(q_3)}{\Gamma(q_1)\Gamma(q_2)\Gamma(1-q_3)} \right)^{\frac{1}{2}} \, . \label{renormalizedstructureconstants}
\end{equation}
For $\delta/k$ and $\epsilon$ small and equal, we find agreement between the structure constants (\ref{overlap}) and (\ref{renormalizedstructureconstants}) in the functional dependence on the charges $q_i$, as well as for the normalization. This is good  check
on the localization argument we used to compute the metric. 
Indeed, since there can be subtleties
in applying localization arguments to non-compact models, as for instance in the case of the elliptic genus \cite{Troost:2010ud,Ashok:2013pya,Murthy:2013mya,Troost:2017fpk}, the check is useful.
It would be interesting to understand the regularization of volume divergences in three-point correlators in non-compact models better still. 
For instance, it would be informative to derive the structure constants directly from $N=2$ Liouville conformal field theory.

\section{The Topological-Anti-Topological Massive Metric}
\label{ttstar}
In this section, we take first steps in exploring the chiral-anti-chiral metric $g_{m \bar n}$ away from the conformal point.
In particular, we will study how it evolves when we add the almost normalizable deformation proportional to $Y^{-1}$. The crux of the section will be to argue
that the latter problem in the non-compact model reduces to a problem in a compact model. In this section, we closely follow 
\cite{Cecotti:1991me}, to which we must refer for more details on the compact case.

\subsection{\texorpdfstring{The $tt^\ast$ Equation}{}}
In \cite{Cecotti:1991me} the dependence of the chiral-anti-chiral metric on deformation parameters $t_i$ of the model
was analyzed and shown to be governed by differential equations named the $tt^\ast$ equations. These are flatness equations for a unitary gauge connection
which can be written as differential equations for the metric $g_{m \bar{n}}$ (viewed as a matrix $g$) \cite{Cecotti:1991me}
\begin{eqnarray}
\bar{\partial}_i(g \partial_j g^{-1}) - [ C_j, g (C_i)^\dagger g^{-1} ] &=& 0
\nonumber \\
\partial_i C_j - \partial_j C_i + [ g (\partial_i g^{-1}), C_j ] - [ g (\partial_j g^{-1}), C_i ] &=& 0 \, 
\end{eqnarray}
where $C_i$ is the matrix that captures the action of the operator coupling to $t_i$ on the Ramond-Ramond vacuum states, or equivalently,
it is the matrix of chiral ring structure constants.
Moreover, from the relation between a basis generated by anti-chiral operators and a basis formed by acting with chiral operators,
as well as from the CPT invariance of the quantum field theory, one obtains a reality condition relating the topological metric $\eta_{mn}$ with the 
topological-anti-topological metric $g_{m \bar n}$  \cite{Cecotti:1991me}
\begin{equation}
\eta^{-1} g (\eta^{-1} g)^\ast = 1 \, .
\label{eq:reality}
\end{equation}

\subsection{A Family of Theories}
In this subsection, we study a very particular one parameter family of non-compact theories for which the $tt^\ast$ equations
are solvable to a considerable extent.

\subsubsection{The Spectrum and The Toda Equation}
The $t t^\ast$ formalism was developed for deformations of the theory, in particular by relevant and normalizable operators \cite{Cecotti:1991me}.
In the following, we will suppose that it also applies to a family of theories obtained by deformation with an almost normalizable relevant operator,
and study to which conclusions this assumption leads. Note that the operator $Y^{-1}$ belongs to the extend chiral ring defined in section \ref{TCFT}.

Consider then the almost normalizable operator $Y^{-1}$, and (drastically) deform the superpotential of the superconformal theory 
with a term proportional to this operator:
\begin{equation}
W_t = \frac{Y^{-k}}{k} - t Y^{-1} \, .
\end{equation}
As in section \ref{TCFT}, we consider the $k-1$ vacua associated to the operators $Y^{-2}, \dots, Y^{-k}$ and,
by using the operator constraint $Y^{-k-1}=t Y^{-2}$, we note that the multiplication by the operator $Y^{-1}$ acts
on these vacua as the matrix
\begin{equation}
C_{t} = \left( \begin{array}{ccccc}
    0 & 1 &  0 &  \dots  & 0 \\
     0 & 0 & 1 & 0 & \dots  \\
     \dots & \dots & \dots & \dots & \dots \\
     0 & \dots & \dots & 0 & 1 \\
     t & 0 & \dots  & \dots & 0
                  \end{array} \right) \, .
                  \label{C}
\end{equation}
To render the $t t^\ast$ equations more concrete, we also use the explicit form of the topological metric.
Our metric reads $\eta_{i,k+2-i}=1$. The reality constraint (\ref{eq:reality}) then becomes 
\begin{equation}
\langle i | i \rangle \langle k+2-i | k+2-i \rangle = 1 
\end{equation}
where we used the fact that the metric $g$ is diagonal in a basis of operators with fixed R-charges.
We can then define the unknowns
\begin{equation}
\varphi_i = \log \langle i | i \rangle 
\end{equation}
for $i=2, 3, \dots, k$. Using the explicit matrix (\ref{C}), corresponding to a sum of positive roots of $A_{k-2}$ as well as $t$ times
the affine root, we then find  that the $t t^\ast$
equation reduces to a set of $\hat{A}_{k-2}$ Toda equations for $\varphi_i$, after a change of variables \cite{Cecotti:1991me}. These reasonings are as in the compact case \cite{Cecotti:1991me},
with the important caveat that we are discussing a family of theories (and not a deformation of a given theory).
To solve the $t t^\ast$ differential equations explicitly, we preliminarily discuss the boundary conditions.

\subsubsection{The Boundary Conditions}
There are two limiting regimes that are under direct perturbative control. Firstly, when the parameter $t$ is zero, we are at the  conformal point, and the metric is equal to the
conformal metric \eqref{eq:conformalmetric}.  This provides one boundary condition.

Secondly, when the parameter $t$ is large, the infrared dynamics is governed by isolated massive vacua, in which one can 
also compute the metric \cite{Cecotti:1991me}. 
The metric  can be presented in a basis of operators which equal
one in one vacuum and zero in all others. In this basis, the metric in the infrared is again
diagonal and equals
\begin{equation}
g_{m \bar{n}}  \approx \frac{\delta_{m\bar{n}}}{|H(Y_n)|} \, , \label{metricinvacua}
\end{equation}
where $H(Y_n)$ is the Hessian evaluated at vacuum $Y_n$. The first (off-diagonal) correction to this classical limit
is given in terms of the instanton/soliton action of the interpolating solution between two vacua:
\begin{equation}
\frac{g_{m\bar{l}}}{(g_{m \bar{m}} g_{l \bar{l}})^{1/2}} \approx e^{i \alpha} (4 \pi z_{ml})^{-1/2} \exp(-2 z_{ml}) \, ,
\label{correctionmetricinvacua}
\end{equation}
where the soliton action for the soliton interpolating between vacua $m$ and $l$  is 
\begin{equation}
z_{ml} =  |W(Y_m)-W(Y_l)| \, ,
\end{equation}
and $\alpha$ is an arbitrary phase. This establishes the second boundary condition.

\subsubsection{A Parallel}
\label{generalargument}
In this subsection, we make a simple parallel with far reaching consequences.
We recall that the most relevant deformation $t X$ of a compact model with Landau-Ginzburg potential
$X^{k_c}/k_c$ gives rise to a $t t^\ast$ equation which is an affine $\hat{A}_{k_c-2}$ Toda equation \cite{Cecotti:1991me}.
This is also the case for the generalized $t t^\ast$ equation for the family of non-compact models at level $k=k_c$ that we study.
 Moreover, the boundary condition in the ultraviolet of the compact model agrees
with the initial condition for the non-compact model, because the respective metrics 
(\ref{eq:conformalmetric}) are obtained from each other by a change of variables.
Since the boundary conditions, the phase symmetries, as well as the differential equations governing the deformed metric
$g_{m \bar{n}}(t)$ agree, the deformed metrics themselves must coincide.

A simple confirmation of this idea is found by observing, as in section \ref{TCFT}, 
that the spectrum of Ramond sector charges in the spectrum for the non-compact model at level $k$ is
$\{ -1/2+1/k,-1/2+2/k,\dots,1/2-1/k \}$. This agrees on the nose with the spectrum of charges
of a compact model with central charge $c^c=3-6/k_c$ at the same value of the level
$k_c=k$. Thus, the Ward identities imply  that the behavior of the metric near the conformal point is also the same, as it must be due
to the more powerful reasoning above.
The metric in the infrared must then also match, when looked at from the right perspective. We'll see how this works out in practice
in a specific example below.

A side remark is that in \cite{Cecotti:1991me}, the central charge of the parent conformal field theory is read off from the behavior of the metric, e.g. near the conformal point.
We claim now that the metric is pertinent to models with differing central charges. These statements are reconciled by the fact that
in \cite{Cecotti:1991me}  a direct relation between the maximal Ramond-Ramond R-charge and the central charge was exploited, which is only valid in a compact model. 
That link is severed in the non-compact model under consideration, once more because the  $SL(2,\mathbb{C})$ invariant vacuum is not part of the normalizable state space.
If one exploits the different connection between the maximal Ramond-Ramond R-charge, the gap between R-charges, and the central charge in the 
non-compact model (namely, $q_{max}+2/k=c/6$), one recovers a consistent
picture. We will illustrate this generic reasoning with the example of the non-compact model at level $k=3$, where we can be very explicit.

\subsubsection{\texorpdfstring{Example: Level $k=3$}{}}
In this subsection, we explicitly determine the metric for a family of superpotentials $W_t^{(3)}$ at level $k=3$
\begin{equation}
W_{t}^{(3)} = \frac{Y^{-3}}{3} - t Y^{-1} \, .
\end{equation}
The classical vacua of the theory are at $Y^{-1}=0,+\sqrt{t}$ and $-\sqrt{t}$. As argued previously, we will consider
the vacua at $0$ to be irrelevant (namely, at an infinite distance in field space, or corresponding to not strictly normalizable states), 
and concentrate on the vacua at $Y^{-1}=\pm \sqrt{t}$. The spectrum of Ramond-Ramond sector normalizable ground state charges is $\{ -1/6, +1/6 \}$.
The solution for the metric must parallel the solution of the compact model at level $k_c=3$, discussed in \cite{Cecotti:1991me}. We follow their discussion
with small but important modifications.

Firstly, we rewrite the $t t^\ast$ equation, still under the crucial assumption that it is valid for the almost normalizable operator proportional to $Y^{-1}$.
Since the metric does not depend on the phase of $t$, we can define $x=|t|^2$ and $y(x)=\langle Y^{-3}|Y^{-3}\rangle$. The reality condition 
\eqref{eq:reality} tells us that also $y(x)^{-1}=\langle Y^{-2}|Y^{-2}\rangle$. Then the $tt^\ast$ equation becomes the Painlev\'e III equation \cite{Cecotti:1991me}
\begin{equation}
\frac{d}{dx}\left(x\frac{d}{dx}\log y\right)=y^2-\frac{x}{y^2} \, . \label{PIII}
\end{equation}
Using the initial condition of the metric \eqref{eq:conformalmetric} at the conformal point, one sets 
\begin{equation}
y^2(t=0)=\frac{\langle Y^{-3}|Y^{-3}\rangle}{\langle Y^{-2}|Y^{-2}\rangle}\bigg|_{t=0}
=3^{\frac{2}{3}}\left( \frac{\Gamma(\frac{2}{3})}{\Gamma(\frac{1}{3})} \right)^2\, . \label{bc}
\end{equation}
These equations already fix the metric, but we can moreover predict the behavior of the solution at large deformation parameter $t$, where the vacua become massive
and well-separated. The behavior at large $t$ is governed by the existence of the two vacua at $Y^{-1}= \pm \sqrt{t}$, and the metric is diagonal in a basis
associated to these vacua (as described around equations (\ref{metricinvacua}) and (\ref{correctionmetricinvacua})) and determined by the Hessian 
\begin{equation}
|H|_{\pm \sqrt{t}} = 2 t^{5/2} \, .
\end{equation}
We note that the Hessian is distinct in the compact and the non-compact case. We will return to discuss this point shortly.
The corrections (\ref{correctionmetricinvacua}) to the metric (\ref{metricinvacua}) are determined by the action of the soliton interpolating between these two vacua
\begin{equation}
z =  | W(\sqrt{t})-W(-\sqrt{t})| = 4/3 |t|^{3/2}=4/3 x^{3/4} \, .
\end{equation}
We still need to link the operators that create the separated vacua with the operators $Y^{-2}$ and $Y^{-3}$ that create the vacua with fixed R-charge in the 
conformal point. To that end, we first define the operators $l_{\pm}$ that take value one on the $\pm \sqrt{t}$ quantum vacua and zero otherwise.
As a consequence of this definition, we must have the operator relations
\begin{eqnarray}
Y^{-2} &=& t (l_++l_-)
\nonumber \\
Y^{-3}  &=& t^{3/2}(l_+-l_-) \, , \label{operatorlink}
\end{eqnarray}
valid at large $t$. These operator relations also differ between the non-compact and the compact model. See \cite{Cecotti:1991me}.

With these definitions for the operators $l_\pm$ that create the separated vacua at large $t$, we have the leading
metric behaviors 
\begin{eqnarray}
\langle l_\pm | l_\pm \rangle &=& \frac{1}{2 |t|^{5/2}}+ \dots
\nonumber \\
\langle l_\pm | l_\mp \rangle &=& \frac{\beta}{2 |t|^{5/2}} z^{-1/2} e^{-2z} + \dots \, , \label{massivevacuametric}
\end{eqnarray}
where $\beta$ is a numerical coefficient.
For the ratio of the norms then, we find the large $t$ asymptotics 
\begin{eqnarray}
\frac{\langle Y^{-3} | Y^{-3} \rangle}{\langle Y^{-2} | Y^{-2} \rangle}
&=& t (1 - 2 \beta z^{-1/2} e^{-2z} + \dots) \, .
\,  \label{asymptotics}
\end{eqnarray}
Crucially, we note that the different asymptotics for the individual norms of the states in the non-compact and the compact model drop out in this ratio.
Secondly, the different behavior (\ref{massivevacuametric}) of the metric near the individual large $t$ vacuum states is  due to the different link (\ref{operatorlink}) between
the operators $l_\pm$ and the conformal field theory operators. 

Finally,  we recall  that the regular solution
to the Painlev\'e III differential equation (\ref{PIII}) which satisfies
the boundary condition (\ref{bc}) indeed has the desired asymptotics (\ref{asymptotics}) \cite{Cecotti:1991me}.
Using the results in \cite{Cecotti:1991me}, one can  compute the value of the numerical constant $ \beta = - 1/(2 \sqrt{\pi}) $.
In short, the metric is identical to the metric in a compact model, as argued in more generality in subsection \ref{generalargument}.

\section{Conclusions}
\label{conclusions}
We twisted a non-compact $N=2$ superconformal field theory with central charge $c=3+6/k$  into a topological conformal field theory 
with a  chiral ring of dimension $k-1$. The original theory, and its topological counterpart admit massive supersymmetry preserving deformations.
We computed the resulting deformed chiral ring, and uncovered the intricate combinatorial solution to the topologically twisted quantum field theories. We moreover showed how a family of such theories,
this time parameterized by the coefficient of an almost-normalizable operator, have a topological-anti-topological metric which is governed by the same $t t^\ast$ equations and boundary conditions
as a deformed compact model at central charge $c=3-6/k$. 

Clearly, this is just the beginning of the exploration of these models. An open problem is to use analytically continued $N=2$ Liouville theory correlators to obtain the chiral and chiral-anti-chiral ring structure constants at the conformal point. Though from the  perspective of this paper this may look trivial, it would be a step forward to understand the localization mechanism directly in terms of conformal field theory. Secondly, we may look for a graphical understanding of the combinatorics inherent in the solution of our models. Thirdly,
we want
to match our solution of the topological quantum field theory to an integrable hierarchy and a Fr\"obenius geometry. Fourthly, we may want to couple the non-compact models to gravity,
and investigate the resulting string theories and putative matrix model duals. Finally, let us mention the possibility to study  our models on Riemann surfaces with boundary. In summary, further effort is 
motivated.

\section*{Acknowledgments}
It is a pleasure to thank our colleagues  for creating a stimulating research environment. We
acknowledge  support from the grant ANR-13-BS05-0001.

\appendix

\section{Examples}
\label{examples}
In this appendix, we provide explicit formulas for the topological quantum field theories that we study in the bulk of the paper, for low levels. 
Firstly, we recall the solutions of the compact models up to level $k_c=5$.
Secondly, we discuss the non-compact models, also up to level $k=5$. We compute the superpotential $W$ by integration from its various derivatives,
and integrate the linear differential equations for the free energy $F$, which is the generator of correlation functions.
\subsection{Compact Examples}
\label{compactexamples}
We start out by describing the compact models. We start from the most trivial model, at level $k_c=2$, and work our way up to level $k_c=5$.
For easy reference
we provide many formulas, and comment little.
\subsubsection*{\texorpdfstring{Level $k_c=2$}{}}
At level two, we have the formulas
\begin{equation}
\phi_0 = 1
\end{equation}
and
\begin{equation}
 \phi_1 = W_X = X \qquad \quad
-\phi_0 =  \partial_0 W = -1 \, ,
\end{equation}
and therefore
\begin{equation}
W = X^2/2-t_0  \, .
\end{equation}
The ring consists of the unit operator only and is undeformed. The structure constants are trivial and follow from the free
energy  
\begin{equation}
F(t_0)=t_0^3/3! \, .
\end{equation}
\subsubsection*{\texorpdfstring{Level $k_c=3$}{}}
At level three, we find
\begin{equation}
\phi_0 = 1 \qquad
\phi_1 = X \qquad
W_X = X^2 - t_1
\, .
\end{equation}
We integrate to
\begin{equation}
W = X^3/3 - t_1 X - t_0 \, .
\end{equation}
The ring is two-dimensional. We can potentially deform by deformation parameters $t_0,t_1$ of R-charge $1,2/3$. The operators in the ring
have R-charge $0,1/3$ and therefore they can not mix (polynomially). 
The function $F(t_0,t_1)$ is equal to
the triple integral of the structure constants and is equal to
\begin{equation}
F(t_0)=t_1 t_0^2/2  +t_1^4/4! \, \, .
\end{equation} 
\subsubsection*{\texorpdfstring{Level $k_c=4$}{}}
In this case, the calculations run
\begin{align}
 \phi_0 &= 1
&
\phi_1 &= X
\nonumber \\
 \phi_2 &= X^2 - t_2 
&
W_X &= X^3 - 2 t_2 X -t_1 
\, ,
\end{align}
as well as
\begin{equation}
W = X^4/4 - t_2 X^2 -t_1 X  +t_2^2/2-t_0 \, .
\end{equation}
 The function $F(t_0,t_1,t_2)$ is
 \begin{equation}
 F(t_0,t_1,t_2)= t_0 t_1^2/2 +t_0^2 t_2/2+  t_2^2 t_1^2/4 +t_2^5/60 
 \, .
 \end{equation}
\subsubsection*{\texorpdfstring{Level $k_c=5$}{}}
The formulas are
\begin{align}
\phi_0 &=1 &  
\phi_1 &= X
\nonumber \\
\phi_2 &= X^2 - t_3 & 
\phi_3 &= X^3 - 2 t_3 X -t_2 
\nonumber \\
W_X &= X^4 - 3 t_3 X^2 - 2t_2 X - t_1 + t_3^2 \, , & &
\qquad  
\end{align}
as well as
\begin{eqnarray}
W &=& X^5/5-t_3 X^3 - t_2 X^2 -t_1 X + t_3^2 X -t_0 + t_2 t_3 \, .
\end{eqnarray}
The free energy works out to be
\begin{equation}
F(t_i) =  t_0 t_1 t_2 + t_0^2 t_3 /2 + t_1^3/3!+ t_1 t_2^2 t_3/2+t_1^2 t_3^2/4 +t_2^4/12+ t_2^2 t_3^3/6 \, .
\end{equation}

\subsection{Non-Compact Examples}
\label{noncompactexamples}
We turn to non-compact examples, and integrate up all the way to the generating function $F$. We list the formulas, and only make a few remarks.
\subsubsection{\texorpdfstring{Level $k=2$}{}}
The model has only one operator, which squares to zero. The superpotential term  demonstrates the non-linearity in the parameter
$s_0$.
\begin{align}
\phi_2 & = s_0 Y^{-2}
&
W_Y &=  -s_0^2 Y^{-3}
\nonumber \\
 W &= s_0^2 \frac{Y^{-2}}{2}
&
F &= 0 \, .
\end{align}
\subsubsection{\texorpdfstring{Level $k=3$}{}}
There are two operators, and the ring is already non-trivially deformed at this level. The generator of correlation functions has a denominator which is a monomial
in the parameter $s_0$:
\begin{align}
\phi_2 &= 
 s_0 Y^{-2}
&
\phi_3 &=
s_0^2 Y^{-3} +s_1 Y^{-2}
\nonumber \\
W_Y &= -s_0^3 Y^{-4} -2 s_1 s_0 Y^{-3}
&
W &= \frac{Y^{-3}}{3} (s_0^3 + 3 s_1 s_0 Y)
\nonumber \\
F & =  
\frac{1}{12} \frac{s_1^4}{s_0^2} \, . 
\end{align}

\subsubsection{\texorpdfstring{Level $k=4$}{}}
The homogeneity properties of the free energy are obvious from the following example
\begin{align}
\phi_2 &= s_0 Y^{-2}
&
\phi_3 &= s_0^2 Y^{-3} + s_1 Y^{-2}
\nonumber \\
\phi_4 &= s_0^3 Y^{-4} + 2 s_0 s_1 Y^{-3} + s_2 Y^{-2}
&
W &= \frac{Y^{-4}}{4} (s_0^4+4 s_0^2 s_1 Y + 2 (s_1^2+ 2 s_0 s_2) Y^2)
\nonumber \\
F &=  \frac{s_1^6-6 s_2 s_1^4 s_0 + 12 s_2^2 s_1^2 s_0^2 -4 s_2^3 s_0^3}{24 s_0^4} \, .
\end{align}

\subsubsection{\texorpdfstring{Level $k=5$}{}}
For good measure, we throw in a final polynomial coding the correlation functions of the level $5$ non-compact model:
\begin{eqnarray}
F &=& \frac{1}{120 s_0^6} \Big(20 s_1^2 s_0^3 \left(3 s_3^2 s_0-4 s_2^3\right)+15 s_2 s_0^4 \left(s_2^3-4 s_3^2 s_0\right)+120 s_3 s_2^2 s_1 s_0^4-120 s_2 s_3 s_1^3 s_0^3
\nonumber \\
&&
+24 s_3 s_1^5 s_0^2+90 s_2^2 s_1^4 s_0^2-36 s_2 s_1^6 s_0+5 s_1^8
\Big) \, .
\end{eqnarray}

\section{A Few Properties of Formal Power Series}
\label{formalpowerseries}
Formal power series are a useful tool in the fields of combinatorics, vertex operator algebras and others. For our purposes, it is useful to 
recall some rules of calculus valid for formal power series $f$ (with finitely many non-zero negative power coefficients):
\begin{equation}
f = \sum_{n \in \mathbb{Z}}^\infty f_n x^n \, .
\end{equation}
We define the bracket operation that picks out certain coefficients of a formal power series\footnote{In the bulk of the paper, we also use the bracket operation to pick a subset
of terms in a formal power series -- we trust the context will provide sufficient guidance to distinguish these two uses.}
\begin{equation}
{[} f {]}_n = f_n
\end{equation}
and has the property that the residue of the derivative of a formal power series is zero
\begin{equation}
Res(f') = [f']_{-1} = 0 \, , 
\end{equation}
which has many non-trivial consequences when cleverly applied.
A formal power series whose constant coefficient $f_0$ is zero and whose linear coefficient $f_1$ is non-zero has a formal inverse. Suppose  that $g$ is the inverse of $f$ (where $f_0=0$ and $f_1 \neq 0$).
We then have the Lagrange inversion formula relating the coefficients of powers of these power series:
\begin{equation}
m [g^n]_m = n [f^{-m}]_{-n} \, .
\end{equation}
For $m=-1$, we find the equation relating the residue of the power of the inverse series $g$ to the coefficients of the original series $f$
\begin{equation}
- [g^n]_{-1} = n [f]_{-n} \, .
\end{equation}
The latter Lagrangian inversion formula is useful in the solution of the compact topological quantum field theory. The solution of the non-compact topological quantum field theory
requires  more advanced techniques, described in the bulk of the paper.

\bibliographystyle{JHEP}

\end{document}